\documentclass[conference]{IEEEtran}
\IEEEoverridecommandlockouts
\usepackage{cite}
\usepackage{amsmath,amssymb,amsfonts}
\usepackage{algorithmic}
\usepackage{graphicx}
\usepackage{textcomp}
\usepackage{xcolor}
\usepackage{siunitx}

\usepackage[utf8]{inputenc} 
\usepackage[T1]{fontenc}    
\usepackage{hyperref}       
\usepackage{url}            
\usepackage{booktabs}       
\usepackage{amsfonts}       
\usepackage{nicefrac}       
\usepackage{microtype}      
\usepackage{lipsum}
\usepackage{bm}
\usepackage{makecell}
\usepackage{graphicx}
\usepackage[ boxed]{algorithm2e}
\usepackage{float}
\usepackage{pgfplots}

\def\BibTeX{{\rm B\kern-.05em{\sc i\kern-.025em b}\kern-.08em
    T\kern-.1667em\lower.7ex\hbox{E}\kern-.125emX}}
\pgfplotsset{compat=1.17}

\begin{document}

\title{Dual perspective method for solving the point in a polygon problem}
\author{\IEEEauthorblockN{Karim M. Ali}
\IEEEauthorblockA{\textit{Aerospace Engineering, Cairo University} \\
12613, Giza, Egypt \\
karim2020@cu.edu.eg}
\and
\IEEEauthorblockN{Amr Guaily}
\IEEEauthorblockA{\textit{Engineering Mathematics and Physics, Cairo University} \\
12613, Giza, Egypt \\
\textit{Smart Engineering Systems Research Center, Nile University} \\
12588, Shaikh Zayed City, Egypt \\
aguaily@nu.edu.eg}
}
\maketitle

\begin{abstract}
A novel method has been introduced to solve a point inclusion in a polygon problem. The method is applicable to convex as well as non-convex polygons which are not self-intersecting. The introduced method is independent of rounding off errors, which gives it a leverage over some methods prone to this problem. A brief summary of the methods used to solve this problem is presented and the introduced method is discussed. The introduced method is compared to other existing methods from the point of view of computational cost. This method was inspired from a Computational Fluid Dynamics (CFD) application using grids not fitted to the simulated objects.
\end{abstract}


\section{Introduction}
The point in a polygon problem is one of the most important problems in computational geometry. The problem poses a question regarding the location of a point in a plane being inside, on, or outside the borders of a polygon. Despite being trivial to a human to solve the problem by sight, automating the problem is not trivial. The problem has applications in a variety of fields including CFD, image processing, cartography \cite{cartograph}, computer vision, etc. In CFD, the level set method \cite{lvlset}, the immersed boundary method \cite{ImmersedBoundaryMethods} and other similar methods depend on the concept of a non-deforming grid that does not adjust to the topology of the simulated object. Rather, the grid- Cartesian for example- can be thought of as a background grid on top of which the simulated object is imposed. In this case, the CFD solver would have to classify the nodes of the grid into solid nodes (inside the polygon) and fluid nodes (outside the polygon).

In the following analysis, it is assumed that the investigated polygon is represented by a set of points that will be named vertices $v_j$. These vertices are connected by a set of edges $e_j$. It is also assumed that the connectivity of these vertices and edges are known. That is for each vertex, the previous and the next vertices are known. Also, the edges connecting these vertices are known. This connectivity is nothing but a generalization to the case where the vertices are listed in an ordered fashion. The vertices and edges connectivity is assumed to be constant. So, if a polygon is deformable, it is assumed that the vertices change their locations in a manner that holds this connectivity constant (i.e. no vertex is allowed to switch indices with another vertex).

One of the earliest methods to solve the point in a polygon problem is the ray casting algorithm \cite{ray} in which a ray emanates from the investigated point to a point that is known for sure to be outside the polygon. The number of times this ray intersects the polygon is recorded. If this number is odd, then the point is located inside the polygon. If it is even, then the point is outside the boundaries of the polygon.

The winding number algorithm \cite{Wesley} is another popular method for solving the point in a polygon problem. It simply depends on constructing a line between the investigated point and each two consecutive vertices in the polygon. The inscribed angle between these two lines is calculated. The process is repeated for all the vertices and the inscribed angles are added. The winding number ($w$) is the number of turns around the investigated point made by sweeping along the polygon. It is simply the summation of the inscribed angles divided by $2\pi$. If the winding number is greater than $0$ then the point is inside the boundaries of the polygon. Otherwise, it is considered as outside. The calculation of the inscribed angles summation is computationally expensive due to the usage of inverse trigonometric functions and is also prone to rounding off errors.

Hormann and Agathos \cite{wind} presented an upgrade to the conventional winding number method. They presented two algorithms named algorithm $6$ (Efficient standard algorithm) and algorithm $7$ (Efficient boundary algorithm). Algorithm $6$ dismisses the problems associated with the conventional winding number method by using only integer changes in the winding number based on some nested if/else statements. Algorithm $7$ remedies some of the special cases not covered in algorithm $6$ regarding the point being on an edge of the polygon by performing extra computations. Thus, algorithm $7$ is more general than algorithm $6$ but slower.   

The sum of areas algorithm \cite{cartograph} is a simple algorithm suitable to convex polygons only. In this method, the investigated point is connected to each two consecutive vertices to form a triangle. The area of this triangle is calculated and the process is repeated for all the edges of the polygon. The summation of these areas is compared to the original area of the polygon. If the calculated area equals the original polygon area then the point is considered inside the polygon. If the areas are not equal, then the point is considered outside the polygon.

El-Salamony and Guaily \cite{salamony} presented the modified polygon method. The method starts by listing the polygon's vertices and then uses this list in computing the original area of the polygon. The idea is that the investigated point can be used to modify the original area of the polygon by including this point as one of the polygon's vertices. The location of the investigated point in the modified list of vertices is determined based on the index of the closest vertex to the point. By comparing the original and modified areas, it can be decided whether the investigated point is inside or outside the polygon boundaries.

The road-map of the following sections starts by presenting the basis on which the presented method is built. Afterwards, a cost comparison between some of the methods mentioned previously will be presented. Finally, a problem related to slender polygons will be discussed and some useful remedies will be suggested. Algorithm (\ref{alg:algAll}) represents a pseudo-code for the presented method.

\begin{table}[t]
\centering
\caption{Point classification based on the status of the closest vertex parent edges}
\begin{tabular}{>{\centering}m{1.2cm}>{\centering}m{3.5cm}>{\centering}m{2.0cm}>{\centering}m{0.5cm}}
\toprule 
Case & Schematic & Rule & Note\tabularnewline
\midrule
\midrule 
$\psi<\pi/2$ & \includegraphics[scale=0.4]{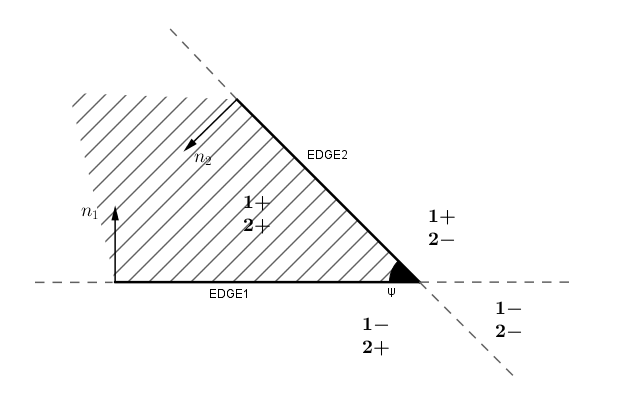} & \makecell{Outside point\\ if both edges \\ agree it's outside} & \makecell{$c>0$}

\tabularnewline
\midrule 

$\psi=\pi/2$ & \includegraphics[scale=0.37]{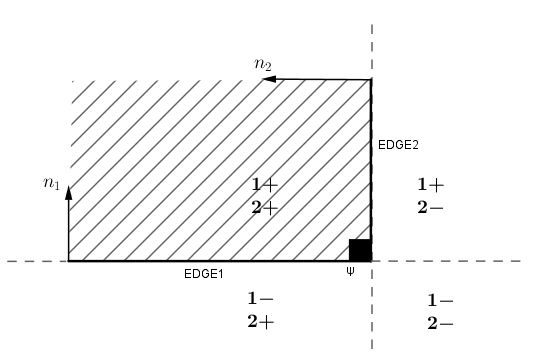} & \makecell{Outside point\\ if both edges \\agree it's outside} & \makecell{$c>0$}

\tabularnewline
\midrule 
$\pi/2<\psi<\pi$ & \includegraphics[scale=0.4]{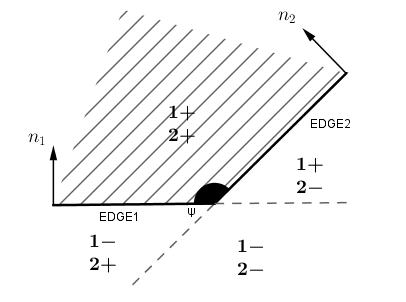} & \makecell{Outside point\\ if both edges \\agree it's outside} & \makecell{$c>0$}

\tabularnewline
\midrule 
$\pi<\psi<3\pi/2$ & \includegraphics[scale=0.40]{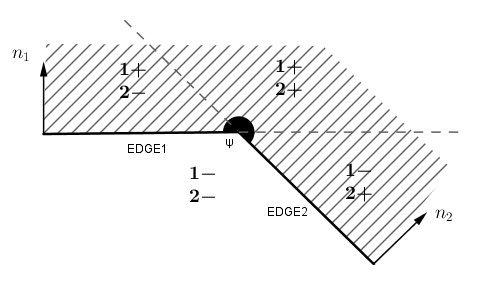} & \makecell{Inside point\\ if both edges\\ agree it's inside} & \makecell{$c<0$}

\tabularnewline
\midrule 
$\psi=3\pi/2$ & \includegraphics[scale=0.37]{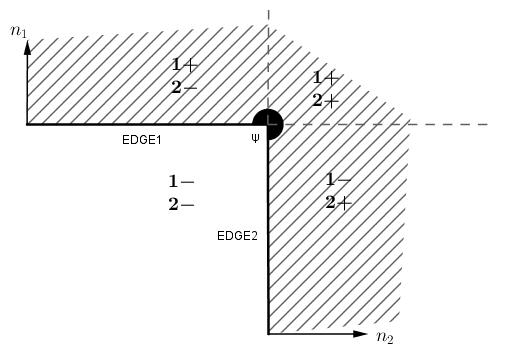} & \makecell{Inside point\\ if both edges\\ agree it's inside} & \makecell{$c<0$}

\tabularnewline
\midrule 

$\psi>3\pi/2$ & \includegraphics[scale=0.35]{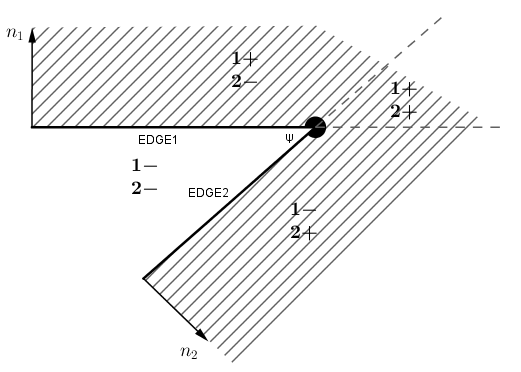} & \makecell{Inside point\\ if both edges \\agree it's inside} & \makecell{$c<0$}

\tabularnewline
\bottomrule
\end{tabular}
\label{tab:class}
\end{table}

\section{The dual perspective method}
A straight line splits a plane into two regions. One of them can be arbitrarily assigned a positive sign and the other a negative sign. A segment can have the same property by fictitiously extending its ends infinitely far away. Thus, any segment can be thought of as a plane split that creates a positive and a negative regions. If we apply the previous analogy by viewing the polygon edges as segments, then each polygon edge splits the plane into two regions, one of them is positive and the other is negative. Since a differentiation between inside and outside regions of the polygon is sought, it is assumed that the region outside the polygon have a positive sign and the one inside have a negative sign. Each polygon edge has its own perspective of what to be considered as inside points and what to be outside points. However, no polygon edge can claim to have the correct perspective. 

Since each vertex is shared by two edges, these two edges can be named the parent edges of the vertex. The edges whose perspectives are considered in the classification problem are the parent edges of the closest vertex to the investigated point. Hence, the first step is to determine the index of the closest vertex to the investigated point and the indices of its parent edges. If the polygon contains $N$ vertices, then this process consumes $\mathcal{O}$($N$) computations. If the coordinates of the investigated point coincides with the closest vertex coordinates, then the investigated point lies on the polygon and the classification process is over.

In table (\ref{tab:class}), schematics of a vertex with its parent edges are presented for different angles between the two parent edges with the true outside region for each case hatched. Each parent edge $e_j$ has an outward pointing normal $\bm{n_j}$. The angle formed by the parent edges is denoted $\psi$. The vector $\bm{A_j}$ is defined to be parallel to edge $e_j$ emanating from the intersection of the parent edges in the direction of the edge $e_j$. The scalar value $c$ is defined as the non-zero component of the resultant vector from the cross product of the vectors $\bm{A_1}$ and $\bm{A_2}$ and its sign is recorded for each row of table \ref{tab:class}.

\begin{equation}
c=(\bm{A_1}\times\bm{A_2}).(0,0,1)^{T}
\label{eqn:cross}
\end{equation}

For each row in the table, each edge in the schematic is extended by a dashed line that splits the plane into two regions. The intersection of these two extensions create four regions each carrying two values with a sign superscript. $1^{+}$ means edge $e_1$ regards this region as an outside region and $1^{-}$ is the opposite. $2^{+}$ and $2^{-}$ have the same definition but regarding edge $e_2$. The straight angle case is absent in table \ref{tab:class} as both parent edges have identical perspectives and any one can be followed.

From table (\ref{tab:class}), whenever the scalar $c$ is positive, the rule is that the point is considered outside only if both parent edges agree it is an outside point. Otherwise, the point is an inside point. Also, whenever the scalar $c$ is negative, the rule is that the point is inside only if both parent edges agree it is inside. Otherwise, the point is outside. Finally, if the scalar $c$ is exactly zero, then a straight angle case is applied.

For any edge $e_j$ to decide the status of a point $p$, a vector $\bm{r_j}$ is constructed form the edge's midpoint $m_j$ to point $p$. The scalar $d$ is defined as the dot product of the edge's outside pointing normal $\bm{n_j}$ and the vector $\bm{r_j}$.
\begin{equation}
d=\bm{n_j}.\bm{r_j}
\label{eqn:side}
\end{equation}

If $d$ is positive, this means that the investigated point lies in the half plane regarded as outside by this edge and vice versa. If $d$ is exactly zero, this means that the point lies either on the edge itself or on the extension of the edge in which case its perspective is dismissed. By this, all the pieces of the dual perspective method have been discussed. Putting these pieces together is the purpose of algorithm (\ref{alg:algAll}). 

After introducing the dual perspective method, it is compared to other existing methods from the point of view of reliability and computational time. The methods to which the dual perspective method is compared comprise the ray casting method, the sum of angles method (the conventional winding number algorithm), algorithms $6$ and $7$ introduced by \cite{wind}, and the modified polygon method introduced by \cite{salamony}.

\section{Results}
\begin{figure}[H]
  \centering
  \includegraphics[width=7cm]{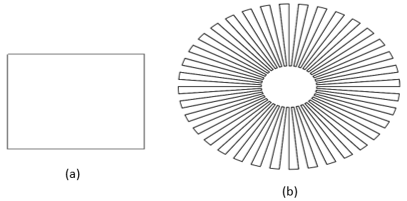}
  \caption{The polygons used in the reliability analysis}
  \label{fig:cases}
\end{figure}

Figure (\ref{fig:cases}) shows two polygons that will be used to test the reliability of the dual perspective method. Shape (a) in figure (\ref{fig:cases}) is a simple $2\times2$ square and defined by only four vertices at its corners. Although it may seem relatively easy, simple shapes can be very tricky in revealing the robustness of an algorithm. Shape (b) of figure (\ref{fig:cases}) comprises a series of five-degree circular sectors spanning radii from $1$ at the inner boundary to $4$ at the outer boundary. The polygon is defined by $97,070$ vertices. 

The testing process of a polygon includes putting the polygon over a $10 \times10 $ square grid discretized into $Q\times Q$ grid points (nodes). Therefore, each testing process comprises solving the point in a polygon problem $Q^2$ times. Testing a few points in the plane is not a guarantee for successful classification of all plane points. However, testing points from all around the polygon would for sure cover all the special case that can accompany the polygon shape. In the following results, the green color symbolizes an outside point. The yellow color symbolizes an inside point, while the black color symbolizes a point that is classified as on-polygon point.

\begin{figure}[H]
  \centering
  \includegraphics[width=9cm]{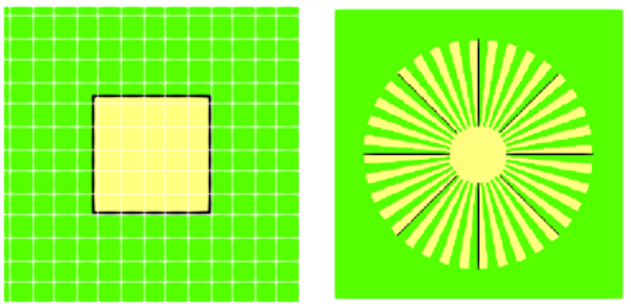}
  \caption{The results of the two test polygons}
  \label{fig:test}
\end{figure}

Figure (\ref{fig:test}) shows the results of classifying $101^2$ grid points for the square polygon case and $40,000$ points for the circular sectors polygon. The method succeeds in classifying all the points correctly for both polygons. Also, the on polygon points are accurately captured. The exact solution of the on polygon points for the square test is straight forward. As for the circular sectors polygon, the background grid consists of vertical lines starting from $x=-5.0$ till $x=5.0$ with a step of $0.05$ creating $201$ vertical lines (considering the background grid as horizontal lines will not change the result). These vertical lines can be represented by

\begin{equation}
x=-5.0+0.05(\alpha-1),
\label{eqn:grideqn}
\end{equation}
where $\alpha$ is an integer ranging from $1$ to $201$.

On the other hand, the polygon consists of two types of curves; segments heading form the inside circle to the outside circle and vice versa, and arcs on two different radii. The segments will be checked for intersections with the grid lines. As previously mentioned, each sector spans five degrees. Thus, each segment exists at an angle which is a  multiple of five degrees. Those segments can be represented by the line equation $\theta=\beta\pi/36$, where $\beta$ is an integer ranging from $0$ to $71$. The segments equation can be re-written in Cartesian coordinates as
\begin{equation}
y=x\tan(\beta\pi/36).
\label{eqn:segeqn}
\end{equation}
The solution of equations (\ref{eqn:grideqn}) and (\ref{eqn:segeqn}) simultaneously yields the points that must be considered as on-polygon points. Substituting equation (\ref{eqn:grideqn}) in equation (\ref{eqn:segeqn}) and solving for $\alpha$ yields
\begin{equation}
\alpha=101+20y/\tan(\beta\pi/36).
\label{eqn:alphaqn}
\end{equation}
By definition, $\alpha$ is an integer. Thus, the fraction in the right hand side should result in a integer value as well. The tangent function in the form $tan(k\pi/r)$ is irrational for rational values of $k$ except for $r=4$ \cite{rational}. Thus, equation (\ref{eqn:alphaqn}) holds only for a multiple of $\pi/4$ as an argument of the tangent function. So, the exact solution of the on-polygon points for this test are the straight lines defined by 
\begin{equation}
y=x\tan(\gamma\pi/4), \gamma=0,1,2...7.
\label{eqn:gamma}
\end{equation}
Since, the solution should exist on the polygon edges, equation (\ref{eqn:gamma}) is bound by the inner and outer radii of the circular sectors polygon.
By checking figure (\ref{fig:test}), the dual perspective method was able to capture all the on-polygon points correctly.

\section{Time consumption}

In this section, the time consumption of the previous methods is compared. The numerical experiments in this section were conducted using a FORTRAN code with an Intel i7-8750H 2.20 GHz processor. As previously mentioned, the used grid has $Q\times Q$ grid points and hence each experiment solves the point in a polygon problem $Q^2$ times. Each experiment is repeated for $z$ times and the total time duration $T_j$ of these $z$ experiments is recorded. To neutralize the effect of the processor state during the running, each $z$ repetitions of an experiment is simulated at ten different times and their total times are averaged. The averaged total time is denoted $T_o$. So, solving $z\times Q^2$ point in a polygon problems takes $T_o$ of time. Thus, the duration of solving one point in a polygon problem $T$ is defined by equation (\ref{eqn:time}).
\begin{equation}
T=\frac{T_o}{z Q^2}=\displaystyle \sum_{j=1}^{10} T_j/ (10z Q^2)
\label{eqn:time}
\end{equation}

\begin{figure}
\centering
\begin{tikzpicture}
\begin{axis}[
    xlabel={$Log \left(N\right)$},
    ylabel={$Log \left( T \left[\mu s\right] \right)$ },
    xmin=0, xmax=5.5,
    ymin=-2, ymax=4,
    xtick={0,1,2,3,4,5},
    ytick={-2,-1,0,1,2,3,4},
    legend pos=north west,
    ymajorgrids=true,
    grid style=dashed,
    width=0.5\textwidth,
    height=0.45\textwidth,
]
 
\addplot[
    color=blue,
    mark=square,
    thick,
    ]
    coordinates {
(0.602059991,-1.346771025)
(1.806179974,-0.899950972)
(2.556302501,-0.277003868)
(2.857332496,0.070425422)
(3.202760687,0.42087397)
(3.380211242,0.667049826)
(3.681241237,1.092910981)
(4,1.404761721)
(4.568201724,2.019059761)
(4.98708503,2.792117534)
(5.164483719,3.041681875)
    };
    \addlegendentry{Dual Perspective}
    
\addplot[
    color=red,
    mark=triangle,
    thick,
    ]
    coordinates {
(0.602059991,-1.175535304)
(1.806179974,-0.191015378)
(2.556302501,0.531569389)
(2.857332496,0.828155883)
(3.202760687,1.179333067)
(3.380211242,1.362194064)
(3.681241237,1.671862751)
(4,1.992610856)
(4.568201724,2.600489928)
(4.98708503,3.186561051)
(5.164483719,3.398311214)
    };
    \addlegendentry{Sum of angles}
    
\addplot[
    color=black,
    mark=*,
    thick,
    ]
    coordinates {
(0.602059991,-1.879315133)
(1.806179974,-0.888377407)
(2.556302501,-0.211245284)
(2.857332496,0.132611063)
(3.202760687,0.483156523)
(3.380211242,0.685413753)
(3.681241237,1.062433596)
(4,1.373402226)
(4.568201724,1.985078238)
(4.98708503,2.809389565)
(5.164483719,3.035892678)
    };
    \addlegendentry{Alg.7}
    
\addplot[
    color=yellow,
    mark=x,
    thick,
    ]
    coordinates {
(0.602059991,-1.946765983)
(1.806179974,-1.002042972)
(2.556302501,-0.352125409)
(2.857332496,0.003667185)
(3.202760687,0.358752066)
(3.380211242,0.607931167)
(3.681241237,1.016535105)
(4,1.303111219)
(4.568201724,1.898404267)
(4.98708503,2.768346666)
(5.164483719,2.987209753)

    };
    \addlegendentry{Alg.6}
    
\addplot[
    color=green,
    mark=o,
    thick,
    ]
    coordinates {
(0.602059991,-1.27750769)
(1.806179974,-0.170738816)
(2.556302501,0.630968235)
(2.857332496,0.925776776)
(3.202760687,1.2948683)
(3.380211242,1.399968177)
(3.681241237,1.697140784)
(4,2.091000107)
(4.568201724,2.652870677)
(4.98708503,3.33114145)
(5.164483719,3.520620891)

    };
    \addlegendentry{Ray casting}
    
\addplot[
    color=brown,
    mark=+,
    thick,
    ]
    coordinates {
(0.602059991,-1.286677726)
(1.806179974,-0.675888801)
(2.556302501,0.005378864)
(2.857332496,0.39340355)
(3.202760687,0.81407681)
(3.380211242,1.025722526)
(3.681241237,1.394692362)
(4,1.700388498)
(4.568201724,2.309979582)
(4.98708503,3.128107689)
(5.164483719,3.374759957)

    };
\addlegendentry{Modified polygon}
 
\end{axis}
\end{tikzpicture}
\caption{Time consumption by each method as a function of the number of polygon vertices}
\label{fig:times}
\end{figure}
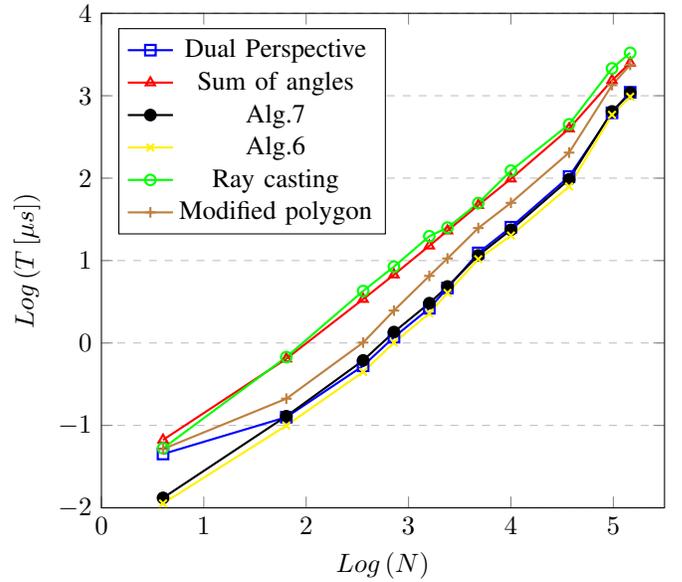

Normally, $T$ has a very small value. So, it is measured in micro-seconds ($10^{-6}$ seconds). Figure (\ref{fig:times}) shows the time consumed by each method to solve one point in a polygon problem as a function of the number of polygon vertices $N$. For adequate scaling, the logarithm to base $10$ of the number of polygon vertices is distributed along the $x$-axis, while the logarithm to base $10$ of the time $T$ expressed in micro-seconds is distributed along the $y$-axis. According to figure (\ref{fig:times}), algorithm ($6$) has the least time consumption in all the conducted experiments. However, algorithm ($6$) has low reliability as it mis-classifies some configurations. On the other hand, the ray casting method and the sum of angles method (conventional winding number algorithm) share the highest time consumption throughout the conducted experiments. In experiments with low number of polygon vertices, the dual perspective method was relatively slow. However, later on, it corrected its route back and had almost the same time consumption of algorithm ($7$). The latter, held the second rank after algorithm ($6$) in many experiments, and ranked third after algorithm ($6$) and the dual perspective method in other experiments. As for the modified polygon method, in experiments of very small or very large number of polygon vertices, the method was as fast as the sum of angles and ray casting methods. Otherwise, the modified polygon was faster than those two methods, but still fell behind the leading three methods by significant values. 

\section{Slender polygon problem}

\begin{figure}[H]
  \centering
  \includegraphics[width=8cm]{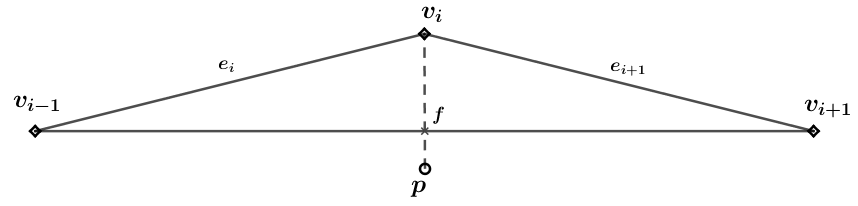}
  \caption{A slender polygon}
  \label{fig:slender}
\end{figure}

El-Salamony and Guaily \cite{salamony2} discussed the problem that arises when the investigated polygon is slender. The problem arises when the path from the closest vertex to the investigated point intersects one of the edges of the polygon. Figure (\ref{fig:slender}) shows a schematic of this case where the closest vertex to the point $p$ is the vertex $v_i$ and the path between them intersects the polygon at the point $f$. In this case, the parent edges become edges $e_i$ and $e_{i+1}$ and both edges classify point $p$ as an inside point. Figure (\ref{fig:slender}) correspond to the case of $\psi>3\pi/2$ in table (\ref{tab:class}), and hence the final classification is that point $p$ is an inside point. Of course, this classification is incorrect. Two remedies can be implemented to avoid this problem.
\begin{itemize}
\item The easiest remedy is to apply a pre-processor to the polygon. The job of the pre-processor is to re-discretize the polygon edges with a length scale that is equal to or less than the length scale of the background grid. This would be more suitable in the case of grids which have a large number of points. As for the case of a single or a few points which do not belong to a grid, the following remedy would be more adequate but is more expensive.

\item To make sure that a case similar to that of figure (\ref{fig:slender}) does not occur, the following should be applied. After the code identifies the closest vertex, the line connecting the closest vertex to the investigated point (the line $v_i p$) is checked for intersection with all the polygon edges. If an intersection occurs with an edge, then the closest vertex is updated with one of the vertices of this edge (the closest to the investigated point) and the dual perspective algorithm is resumed normally. In case of intersection with multiple polygon edges, the intersection points are calculated and the chosen edge would be the one whose intersection point is the closest to the investigated point. This check adds $\mathcal{O}$($N$) more computations to the dual perspective method. To save computational time, this check should be set as optional in the code. It is more appropriate to activate it in case high suspicions of slender polygon(s) existence. For automatic activation of this check, a pre-processor can be implemented to measure the aspect ratio of the investigated polygon(s) and activate the slender polygon check based on a pre-defined tolerance criteria. However, such slender polygons are seldom used in CFD applications due to the numerical instabilities they initiate. 

\end{itemize}

\section{Summary}
The dual perspective method has been discussed, and its pillar stones have been introduced. The problem was posed so that the polygon was defined by a set of vertices interconnected by edges. Each polygon edge can view the relative location of the investigated point from its own perspective. The dual perspective method depended on merging the perspectives of two polygon edges to correctly classify the investigated point. The method showed high robustness in dealing with irregular shapes and sharp edges. The speed of the method was compared to the other methods and the results showed that the dual perspective method is more efficient in the case of moderate and high number of polygon vertices. Finally, some remedies to the slender polygon problem were discussed. 
 \begin{algorithm}
\SetAlgoNoLine
\DontPrintSemicolon
\SetKwInOut{Input}{Input}
\Input{ A polygon defined by $N$ vertices, and a point $p$}
\KwResult{ A decision about point inclusion}
\BlankLine
- Identify the closest vertex $v_c$ to the investigated node $p$ and its parent edges $e_1$ and $e_2$.\;
- Construct the vectors $\bm{A_1}$ and $\bm{A_2}$ using:\;
$\bm{A_j} \longleftarrow$ \emph{vector} ($v_c$, $v_j$).\;
- Compute the scalar $c$ according to equation (\ref{eqn:cross}).\;
- For each parent edge $e_j$ compute the corresponding scalar $d_j$ according to equation (\ref{eqn:side}).\;
- For each parent edge $e_j$ set the flags $s_j$ and $L_j$ according to:\;
\uIf{$d_j > 0$}{
    $s_j=1$\;
  }
  \uElseIf{$d_j < 0$}{
    $s_j=-1$\;
  }
  \Else{
   $s_j=0$\;
    \eIf{point on edge $e_j$}{
    $L_j=1$\;
    }
    {
    // Dismiss the perspective of edge $e_j$.\;
    }
  }
- If an edge is dismissed, follow the perspective of the other parent edge and \underline{exit the code}.\;
- For any of the parent edges $e_j$:\;
\uIf{($s_j=0$ AND $L_j=1$)}{
    The point lies ON the polygon; \underline{Return};\;
  }
- Form a final conclusion according to:\;
\uIf{($c>0$)}{
    \uIf{($s_1=1$ AND $s_2=1$)}{
    The point lies OUTSIDE the polygon\;}
    \Else{The point lies INSIDE the polygon\;}
    }
\uElseIf{($c=0$)}{
    \uIf{($s_1=1$)}{
    The point lies OUTSIDE the polygon\;}
    \Else{The point lies INSIDE the polygon\;}
    }
\Else{
    \uIf{($s_1=-1$ AND $s_2=-1$)}{The point lies INSIDE the polygon\;}
    \Else{The point lies OUTSIDE the polygon\;}
    }
    
\SetKwProg{Fn}{Function}{ is}{end}
\Fn{vector($\bm{a}$, $\bm{b}$)}{return $(\bm{b}_x-\bm{a}_x, \bm{b}_y-\bm{a}_y)$\;}

 \caption{The Dual perspective method}
 \label{alg:algAll}
\end{algorithm}

\end{document}